\documentclass[a4paper,12pt, epsfig]{article}
\usepackage{graphicx}
\usepackage{enumitem}

\newskip\humongous \humongous=0pt plus 1000pt minus 1000pt

\newif\ifdtup

\def\,{\hspace{-.1cm}}

\def\fc#1#2 {\frac{n}{q}#1\frac{n}{q}#2}

\renewcommand{\theequation}{\arabic{section}.\arabic{equation}}
\renewcommand{\(}{\begin{equation}}
\renewcommand{\)}{end{equation} \vspace{-.05in}\linebreak}

\newcounter{saveeqn}
\newcounter{savealpheqn}

\newcommand{\alpheqn}{\setcounter{saveeqn}{\value{equation}}%
  \stepcounter{saveeqn}\setcounter{equation}{0}%
  \renewcommand{\theequation}{\mbox{\arabic{section}.\arabic{saveeqn}
\alph{equation}}}
  \renewcommand{\)}{\end{equation}}}
\def\part#1{\frac{\partial}{\partial{#1}}}%
\def\group#1{\refstepcounter{equation}\setcounter{saveeqn}
 {\value{equation}}%
  \label{#1}\setcounter{equation}{0}%
\renewcommand{\theequation}{\mbox{\arabic{section}.\arabic{saveeqn}
\alph{equation}}}
  \renewcommand{\)}{\end{equation}}}
\newcommand{\reseteqn}{\setcounter{equation}{\value{saveeqn}}%
  \renewcommand{\theequation}{\arabic{section}.\arabic{equation}}%
  \renewcommand{\)}{\end{equation}}}

\newcommand{\aalpheqn}{\setcounter{saveeqn}{\value{equation}}%
  \stepcounter{saveeqn}\setcounter{equation}{0}%
  \renewcommand{\theequation}{\mbox{
        \Alph{subsection}.\arabic{saveeqn}\alph{equation}}}
   \renewcommand{\)}{\end{equation}}}
\newcommand{\areseteqn}{\setcounter{equation}{\value{saveeqn}}%
  \renewcommand{\theequation}{\Alph{subsection}.\arabic{equation}}%
  \renewcommand{\)}{\end{equation}}}

\renewcommand{\thefootnote}{\alph{footnote}}
\renewcommand{\(}{\begin{equation}}
\renewcommand{\)}{\end{equation}}
\newcommand{\ba}{\begin{eqnarray}}
\newcommand{\ea}{\end{eqnarray}}
\newcommand{\cbp}{\mathop{\vtop{\ialign{##\crcr
   $\hfil\displaystyle{}\hfil$\crcr\noalign{\kern-13pt\nointerlineskip}
   \BIG{)}\hskip0pt\crcr\noalign{\kern3pt}}}}}
\newcommand{\pa}{\mathop{\vtop{\ialign{##\crcr
    
$\hfil\displaystyle{\oplus}\hfil$\crcr\noalign{\kern+1pt\nointerlineskip 
}
   \hspace{.08in}$^{\alpha=0}$\hskip6pt\crcr\noalign{\kern3pt}}}}}

\catcode`\@=11
\def\vereq#1#2{\lower3pt\vbox{\baselineskip1.5pt \lineskip1.5pt
\ialign{$\m@th#1\hfill##\hfil$\crcr#2\crcr\sim\crcr}}}
\catcode`\@=12

\renewcommand{\(}{\begin{equation}}
\renewcommand{\)}{\end{equation}}

\newcommand{\beas}{\begin{eqnarray*}}
\newcommand{\eeas}{\end{eqnarray*}}

\newcommand{\bquo}{\begin{quote}}
\newcommand{\enqu}{\end{quote}}

\newcommand{\beq}{\begin{equation}}
\newcommand{\eeq}{\end{equation}}
\newcommand{\bea}{\begin{eqnarray}}
\newcommand{\eea}{\end{eqnarray}}

\newskip\humongous \humongous=0pt plus 1000pt minus 1000pt

\newif\ifdtup

\catcode`\@=11

\@addtoreset{equation}{section}

\def\@normalsize{\@setsize\normalsize{15pt}\xiipt\@xiipt
\abovedisplayskip 14pt plus3pt minus3pt%
\belowdisplayskip \abovedisplayskip
\abovedisplayshortskip \z@ plus3pt%
\belowdisplayshortskip 7pt plus3.5pt minus0pt}

\def\small{\@setsize\small{13.6pt}\xipt\@xipt
\abovedisplayskip 13pt plus3pt minus3pt%
\belowdisplayskip \abovedisplayskip
\abovedisplayshortskip \z@ plus3pt%
\belowdisplayshortskip 7pt plus3.5pt minus0pt
\def\@listi{\parsep 4.5pt plus 2pt minus 1pt
      \itemsep \parsep
      \topsep 9pt plus 3pt minus 3pt}}

\relax

\catcode`@=12

\catcode`\@=11

\def\section{\@startsection{section}{1}{\z@}{3.5ex plus 1ex minus  .2ex}{2.3ex plus .2ex}{\large\bf}}

\def\thesection{\arabic{section}}
\def\thesubsection{\arabic{section}.\arabic{subsection}}

\def\appendix{\setcounter{section}{0}
 \def\thesection{Appendix \Alph{section}}
 \def\thesubsection{\Alph{section}.\arabic{subsection}}
 \def\theequation{\Alph{section}.\arabic{equation}}}
\renewcommand{\theequation}{\arabic{section}.\arabic{equation}}

\begin{document}
\def\thefootnote{\fnsymbol{footnote}}
\def\thetitle{Uncertainty in the Reactor Neutrino Spectrum and Mass Hierarchy Determination}
\def\autone{E. Ciuffoli}
\def\affa{Institute of Modern Physics, NanChangLu 509, Lanzhou 730000, China}
\def\affb{University of the Chinese Academy of Sciences, YuQuanLu 19A, Beijing 100049, China}

\begin{center}
{\large {\bf \thetitle}}

\bigskip

\bigskip

{\large \noindent  \autone{${}^{1}$}\footnote{emilio@impcas.ac.cn}, J. Evslin{${}^{1,2}$}\footnote{jarah@impcas.ac.cn} and H. Mohammed{${}^{2,1}$}\footnote{hosam@impcas.ac.cn}}

\vskip.7cm

1) \affa\\
2) \affb\\

\end{center}

\begin{abstract}
\noindent
One of the challenges that must be overcome in order to determine the neutrino mass hierarchy using reactor neutrinos is the theoretical uncertainty in the unoscillated reactor neutrino spectrum: this is one of the reasons why, recently, it was proposed to add a near detector to the JUNO experiment. A model-independent treatment of the spectrum uncertainty will be discussed, as well as the effect that it will have on the final result. Moreover, since the neutrino spectrum depends on the chemical composition of the fuel, the spectra at the near and far detectors will be different, because they will receive neutrinos from different cores. Taking into account the time evolution of the chemical composition of the fuel in the reactor core, it is possible to reconstruct the far detector spectrum from the near detector data.  We will show that the method used to reconstruct the spectrum can affect sensitivity to the mass hierarchy, however if the near detector is large enough the difference will be negligible.

\end{abstract}

\setcounter{footnote}{0}
\renewcommand{\thefootnote}{\arabic{footnote}}

\section{Introduction}


In the next decade several experiments will attempt to measure the neutrino mass hierarchy \cite{T2K,NOVA,ORCA,PINGU,Dune,JUNO}: some experiments are already taking data \cite{T2Krecent, NOVArecent}, and preliminary results seems to point towards the normal hierarchy \cite{GlobalFit}, however we are far from a clear determination. In reactor neutrino experiments, like JUNO \cite{JUNO}, the mass hierarchy is determined by studying the interference between the 1-3 and 2-3 oscillations at intermediate baselines. There are several challenges that must be overcome in this kind of experiment: an excellent energy resolution is needed ($\Delta E/E\leq 3\%/\sqrt{E}$, where the energy is expressed in MeV), and the systematic errors on the energy reconstruction must be strongly constrained. 

Another problem that must be faced is the uncertainty in the reactor neutrino spectrum: a few years ago, with the measurement of the ``5 MeV bump'', it became clear that the theoretical models for reactor neutrinos are not in agreement with the experimental data; moreover all the experimental data currently available are obtained with an energy resolution considerably worse than the one required for the mass hierarchy determination (at Daya Bay, $\Delta E/E\simeq 7\%/ \sqrt{E}$). It is possible (and there are also some hints in this sense also from {\it ab initio} calculations) that the reactor neutrino spectrum is not smooth as it is assumed to be in the current theoretical models, but there is a ``fine structure'', present but currently undetected, that could affect the determination of the mass hierarchy. This is one of the reasons why, recently, it was announced that the JUNO experiment will have a near detector, called JUNO-TAO (Taishan Antineutrino Observatory) \cite{JUNO-Neutrino,JUNO-NuPhys}, with mass of the order of tons. In this paper we will discuss how the uncertainty on the theoretical spectrum (constrained by the data from the near detector) can affect the mass hierarchy determination. 

An additional complication arises from the fact that the JUNO far detector will receive neutrinos from two different power plants, Taishan (2 cores) and Yangjiang (6 cores), while the near detector will be able to measure the spectrum only of one of Taishan's cores; since the reactors in the two complexes are of different model and generation (EPR, Gen. III at Taishan, CPR-1000, Gen. II at Yangjiang), there will be a difference between the unoscillated spectra at the near and far detector. We will discuss how to parametrize the spectra in this case and, assuming that the reactor neutrino spectrum depends only on the chemical composition of the fuel, we will show that from the time evolution of the near detector spectrum it is possible to reconstruct a spectrum with a generic chemical composition. Some of the results in this paper were also presented at the PhysStat-$\nu$ meeting at CERN, in 2019 \cite{MioPhyStat}.

\section{Mass Hierarchy From Reactor Neutrinos}
More than 15 years ago Petcov and Piai \cite{Petcov} pointed out that it is possible to determine the neutrino mass hierarchy from reactor neutrinos, by studying the oscillation probability at intermediate baselines. The main problem (which is somehow related to most if not all the challenges that must be faced in this kind of experiment) is that there is a strong degeneracy between a change of hierarchy and a shift in $\Delta m_{32}^2$. 
\begin{figure}\begin{center}
\includegraphics[width=0.45\textwidth]{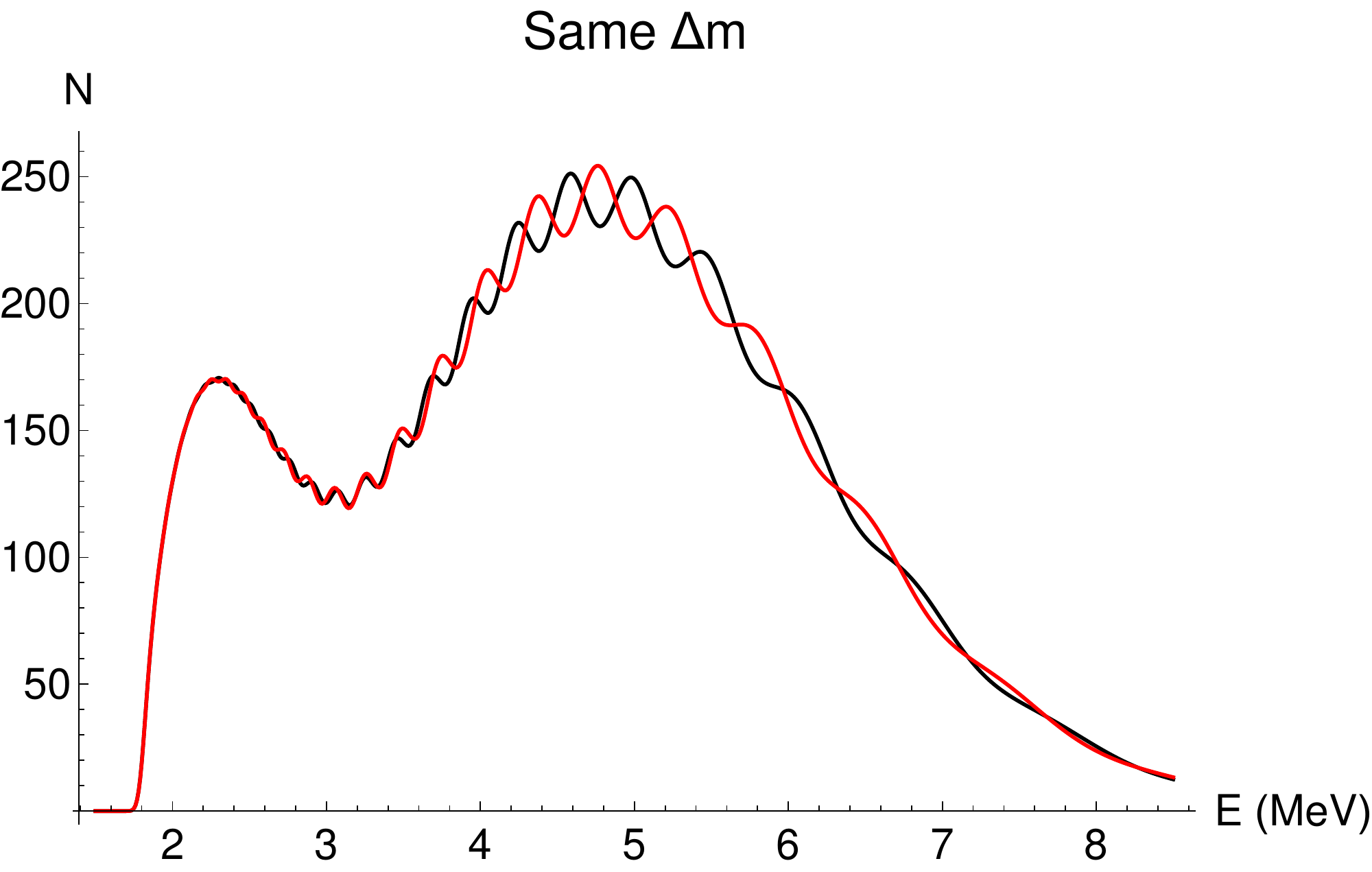}
\includegraphics[width=0.45\textwidth]{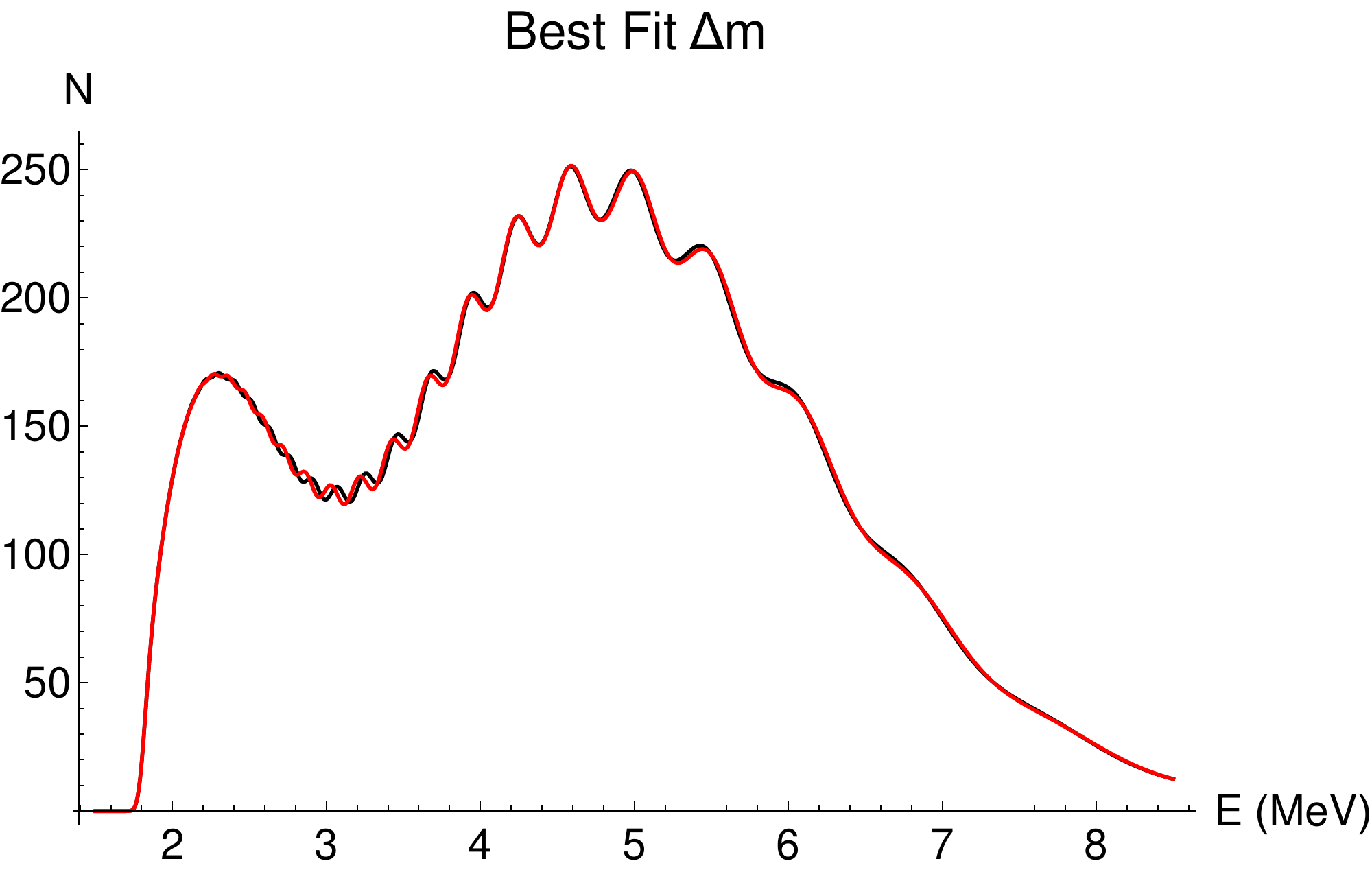}
\end{center}
\caption{\label{SpMH} Expected spectra for normal and inverted hierarchy at 53 km from the source. Left Panel: same value of $\Delta m_{32}^2$ used for both hierarchy. Right Panel: for inverted hierarchy, $\Delta m_{32}^2$ is shifted by $\sim1.5\sigma$'s}
\end{figure}
Indeed, as is shown in Fig. \ref{SpMH}, if we try to fit the normal hierarchy spectrum assuming the inverted hierarchy, it is possible to obtain the same high-energy oscillations by just shifting the best fit value for the inverted hierarchy by less than 1.5 $\sigma$'s (using the best fit values and the precision reported in \cite{PDG}); it should be underlined that, since usually experiments cannot measure directly $\Delta m_{32}^2$ or $\Delta m_{31}^2$ but only an effective mass which depends on the hierarchy, from global fits we get two different values for $\Delta m_{32}^2$, one for each hierarchy; this makes more difficult to break this degeneracy, even if constrains from other experiments are taken into account. 
\begin{figure}\begin{center}
\includegraphics[width=0.45\textwidth]{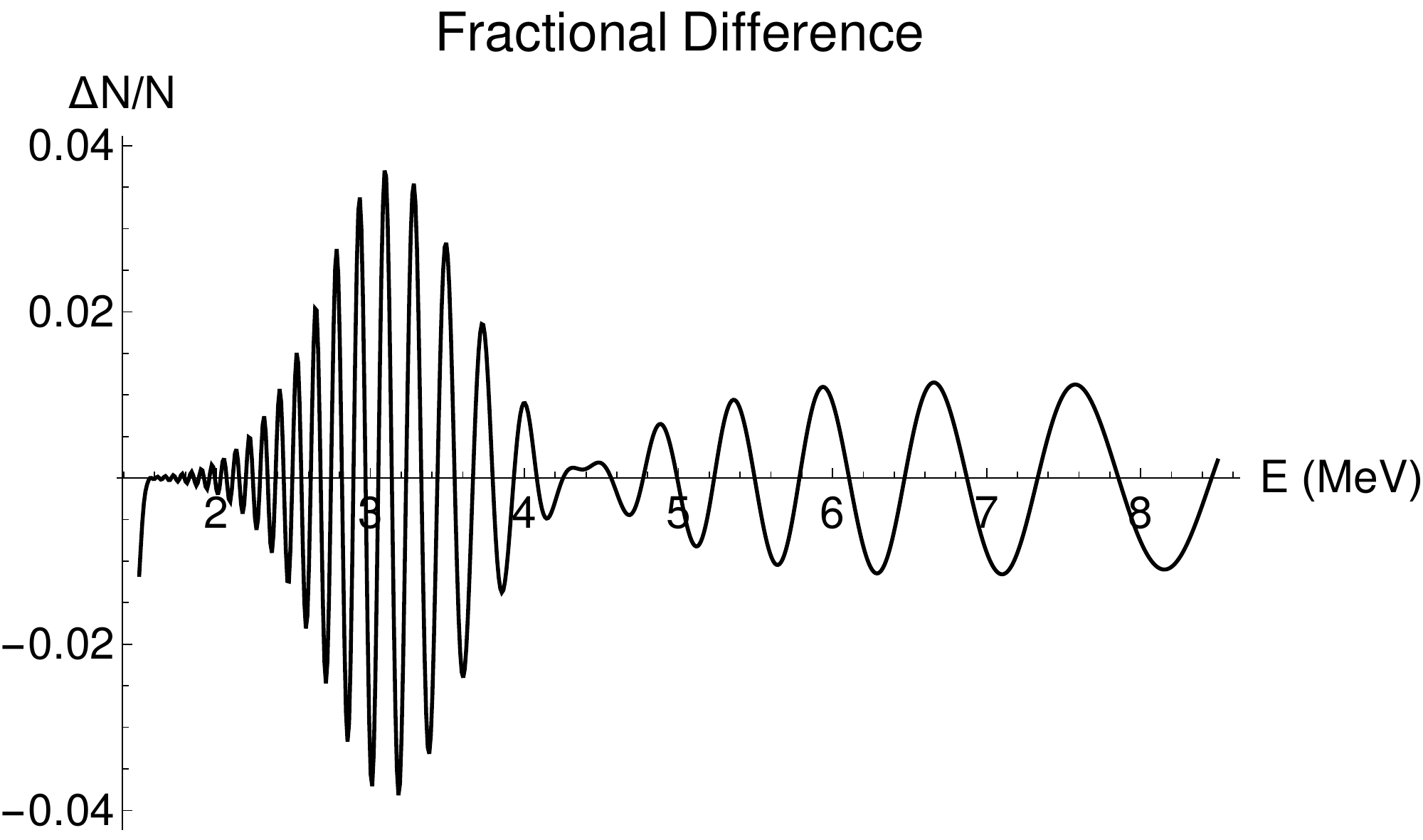}
\includegraphics[width=0.45\textwidth]{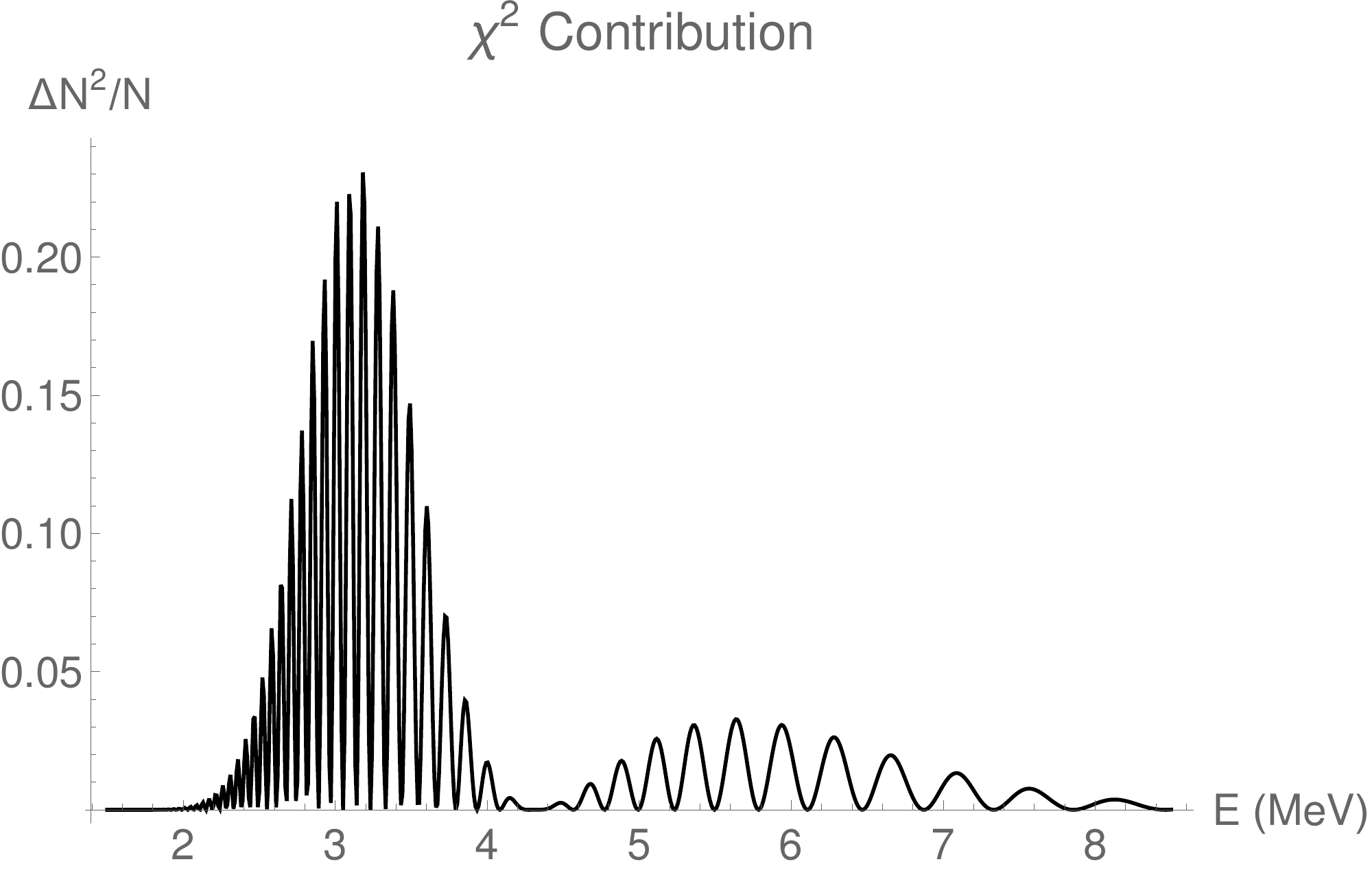}
\end{center}
\caption{\label{DiffSp} Difference between the spectra from the right panel of Fig. \ref{SpMH}.  Left Panel: Fractional difference. Right Panel: $\chi^2$ contribution ({\it i.e.} $\Delta N^2/N)$ }
\end{figure}
The main consequences of the degeneracy can be summarized as follows:
\begin{itemize}
\item Using a $\chi^2$ test the high-energy part of the spectrum is needed to break the degeneracy; the hierarchy can then be determined by comparing the oscillations at low-energy
\item  However, since these oscillation are very fast, an excellent energy resolution is needed at low energies: the goal for JUNO is to achieve an energy resolution of $3\%/\sqrt{E}$; even with such resolution, the energy smearing will significantly decrease the amplitude of the oscillations. From Fig. \ref{DiffSp} it can be read that, using the value of $\Delta m_{32}^2$ that would minimize the $\chi^2$, the difference between the spectra is of the order of $1-4\%$ and that most of the information on the hierarchy comes from the region between 2 and 4 MeV. 
\item Since the hierarchy is determined by a small shift in the position of the low-energy peaks, even a very small systematic error on the energy reconstruction could significantly affect the final result. For this reason, this kind of experiment requires also very good constraints on the non-linearity \cite{mioInterferenza,QianNonLin,CapozziNonLin}.
\end{itemize}

\subsection{Fine Structure}
In nuclear reactors neutrinos are not produced in the fissions, but from the beta decay of the unstable isotopes created. Even if, in theory, it is possible to calculate explicitly the neutrino spectrum from first principles ({\it ab initio} calculations),  it is not yet possible to reach a good precision in this way, due to the large number of isotopes and different decays that must be taken into account (of the order of $10^3$ and $10^4$, respectively).   The flux model most commonly used in reactor neutrino experiments is the Huber+Muller model \cite{Huber, Muller}, where the neutrino spectrum is obtained using the conversion method, starting from the beta spectrum of the isotopes created in the fission process.   Its predictions are however in tension with the measurement of the ``5 MeV bump.''  

Even if it's not possible to calculate precisely the reactor neutrino spectrum from {\it ab initio} calculations, it is nonetheless possible to obtain some important information from this approach. Due to the Coulomb enhancement of low-energy electrons in beta decay, there is a sharp cut-off at the end-point of the energy spectrum of neutrinos from beta decay.  Since, in nuclear reactors, the total neutrino spectrum is given by the sum of a very large number different beta decays, this leads to the presence of sawtooth-like features in the spectrum (also called ``fine structure'') \cite{FineStructDwyer,FineStructHayes}. The fact that these have not yet been observed in reactor neutrino experiments is not surprising, since all the experimental data currently available are obtained with an energy resolution considerably lower than what required for the mass hierarchy determination. Indeed, the fine structure is expected to be very small, of the order of 1-2\%, and such small fluctuations would be completely suppressed by the finite energy resolution at past reactor neutrino experiments.   

However, due to the degeneracy shown in Fig. \ref{SpMH}, even a small uncertainty on the spectrum can affect the sensitivity to the mass hierarchy \cite{Forero}. It was argued in \cite{Fourier,HayesNeutrino} that, using the Fourier transform analysis, the fine structure will have very little effect on the mass hierarchy determination. The reason for this is that the Fourier transform is sensitive only to the global oscillation behavior: the fine structure, in order to affect the determination of the hierarchy, should have a frequency similar to the one of 1-3 and 2-3 oscillations, which the authors claim is unlikely, since the two processes are based on completely different physical mechanisms. Clearly the impact of each systematic error depends on the analysis method and so 
even if the fine structure has less effect on the Fourier transform analysis this does not imply that it can be neglected using, for example, a $\chi^2$ analysis. 

It is also important to notice that there are other serious problems related to the determination of the neutrino mass hierarchy using the Fourier transform. As mentioned before, one of the most difficult challenges in this kind of experiment is to control the systematic errors due to the non-linearity: with a $\chi^2$ analysis, usually a precision of at least $1\%$ is required (which is not trivial to achieve in such a large detector); however using the Fourier transform an even better precision should be required \cite{QianNonLin}; in \cite{Fourier} it is mentioned that for this kind of analysis the precision on the unknown non-linearity should be at least $0.5\%$. Moreover the Fourier transform is more sensitive to the high-energy part of the spectrum, where the flux is very low and the background is more relevant. Using a $\chi^2$ analysis it is possible to simply ignore the high-energy events, since they carry very little information on the mass hierarchy, however if we use the Fourier analysis, a high-energy cut will introduce a spurious and unwanted dependence on $\Delta m_{32}^2$, as was shown in \cite{mioNuisance}, introducing a new source of error. These are some of the reasons why, in the last years, this method was used very rarely in the literature, and in almost all the papers on the topic a $\chi^2$ analysis was used.

\section{$\chi^2$ Analysis}

Our test statistic is defined as:
\begin{equation}
\Delta\chi^2=\chi^2_{IH}-\chi^2_{NH}
\end{equation}
Here $\chi^2_{IH(NH)}$ is the $\chi^2$ statistic where the inverted (normal) hierarchy is assumed. We can write
\begin{equation}\label{ChiTot}
\chi^2_{MH}=\min_{\vec{\alpha}}\left(\chi^2_{MH,F}(\vec{\alpha})+\chi^2_{MH,N}(\vec{\alpha})+P.T.(\vec{\alpha})\right)
\end{equation}
where $MH=NH,IH$, $\vec{\alpha}$ is a vector that contains all the pull parameters (if present) that must be minimized, $\chi^2_{MH,F(N)}$ is the contribution to the $\chi^2$ of the far (near) detector and $PT(\vec{\alpha})$ represents the eventual penalty terms related to the pull parameters. In this paper we will introduce one pull parameter for each energy bin, called $\beta_i$, to parametrize the uncertainty on the reactor neutrino spectrum: there will be no correlation between two of these parameters from different energy bins however, for a given bin, the pull parameter used at the near and far detector will be the same.  Only one other pull parameter will be used, to take into account the uncertainty on the value of $\Delta m_{32}^2$, since the uncertainties over the other mixing parameters, the total flux renormalization, etc... have very little effect on the final result. Even if we introduce one pull parameter for each energy bin, the number of degrees of freedom is still positive, because now we have two data points per bin, one for each detector.

For $\beta_i$ no penalty term will be introduced.  This means that present theoretical and experimental knowledge of the spectra is not used, as existing constraints are negligible compared with those that will come from the near detector's data.  $\sigma_m$ will be $0.05\times 10^{-3}$ eV$^2$, however it should be remembered that the best fit values for $\Delta m_{32}$ are different if the normal or inverted hierarchy is assumed, so this will also affect the penalty term.  
The mass hierarchy is assumed to be normal, we use the following values for the mixing parameters (all the values are taken from the PDG \cite{PDG}, except for $\textrm{sin}^22\theta_{23}$, which is assumed to be 1):
\begin{eqnarray} 
&&\Delta m_{21}^2=7.53\times 10^{-5} \textrm{eV}^2 \qquad \Delta m_{32,NH}^2=2.51\times 10^{-3} \textrm{eV}^2 
\nonumber\\ &&\Delta m_{32,IH}^2=-2.56\times 10^{-3} \textrm{eV}^2 \qquad \sigma_m =0.05 \times 10^{-3} \textrm{eV}^2
\nonumber \\
&& \textrm{sin}^22\theta_{23}=1  \qquad \textrm{sin}^22\theta_{13}=0.083 \qquad \textrm{sin}^22\theta_{12}=0.851 \nonumber
\end{eqnarray}
For the far detector, we will consider an effective mass of 20 ktons, while for the near detector, if not otherwise specified, we consider 3 tons, similar to the mass of JUNO-TAO as reported in \cite{JUNO-Neutrino}. 

It should be noted that, even if we choose parameters similar to the ones of the JUNO experiment, the focus of this study is to investigate the effect of uncertainty in the reactor neutrino spectrum on the mass hierarchy determination, not to discuss the precision that can be achieved in a specific experiment. For this reason, in order to simplify the calculations, we do not take into account effects that are not related with this problem, even if they can strongly affect the expected $\Delta\chi^2$; for example, we assume that all the cores are at the same distance from the far detector, while in reality this is not the case (and this yields a sizeable correction to the final result \cite{mioInterferenzaPrimo}).  Moreover, we do not consider the systematic error related to the non-linearity, which will also strongly affect the sensitivity of the experiment.  All these effects are not related to the topic discussed here, so they will be ignored.  As a result the $\Delta\chi^2$ reported are significantly overestimated and should not be taken as representative of the precision that can be achieved at JUNO.

The expression for $\chi^2_{MH,D}$ (where $D=N,F$) is:
\begin{equation}\label{ChiD}
\chi^2_{MH,D}(\vec{\alpha})=\sum_i\frac{(N_{exp,D,i}-N_{Th,D,i}(MH,\Delta m_{32}^2,\beta_i))^2}{N_{exp,D,i}}+P.T.
\end{equation}
where the index $i$ indicates the energy bin, and runs from 1 to $N_E$ and $P.T.$ is the penalty term for $\Delta m_{32}^2$, namely
\begin{equation}
P.T.=\frac{(\Delta m_{32}^2-\Delta m_{32,IH}^2)^2}{\sigma_m^2}
\end{equation}
In this paper we will use the Asimov data set, where the expected value of $\Delta\chi^2$ for a given experiment is computed using the expected number of events instead of $N_{exp,D,i}$. As a consequence, $\chi^2_{NH}=0$ (since we assumed the hierarchy to be normal) and $\Delta\chi^2=\chi^2_{IH}$. For this reason, from here on we will drop the subscripts $exp$ and $Th$: the two sets can be recognized because the Asimov data set will be calculated assuming the normal hierarchy, the fit assuming the inverted. It should be noticed that, in principle, one could use two different spectra for the fit and for the Asimov data set, because we cannot assume the model we use is the correct one, however since there is a pull parameter for every bin, the difference can be eliminated just by a shift of $\beta_i$; which means that we can safely use the same unoscillated spectrum for both datasets.

The expected number of events at the far detector is given by
\begin{equation} \label{ExpectedEventAnalytical}
N_{MH,F,i}=\int_{E_i}^{E_{i+1}}\textrm{d}E\int \textrm{d}E' G(E;E',\sigma^2(E'))  P_{MH}(E',L) \sum_K \phi_K(E')
\end{equation}
where $K$ indicates the different reactor cores and it is summed over all of them, $E$ is the visible energy, $E'$ the real energy of the neutrino, $G(E;E',\sigma^2(E'))$ is a Gaussian distribution with mean $E'$ and standard deviation $\sigma(E')$ that describe the finite energy resolution of the detector, $\sigma(E')=3/100\sqrt{E'}$, $L$ is the baseline, $P(E',L)$ is the oscillation probability $\phi_K(E')$ the unoscillated neutrino spectrum (taking into account the emission spectrum, the chemical composition of the fuel at the core, the cross section, the detector mass and the eventual detector efficiency; in principle $\phi_K$ depends also from the baseline, due to the geometrical factor, however this dependence is left implicit). Here we assumed that all the cores are at the same distance from the detector; this is not necessarily true, however the generalization is quite straightforward, and it will be briefly discussed in the next section. 

Since, due to the degeneracy with $\Delta m_{32}^2$, the mass hierarchy must be determined by studying relatively small differences between very fast oscillations, the dimension of the energy bins must be small, otherwise these differences would be averaged out by the size of the bin. 
\begin{figure}\begin{center}
\includegraphics[width=0.45\textwidth]{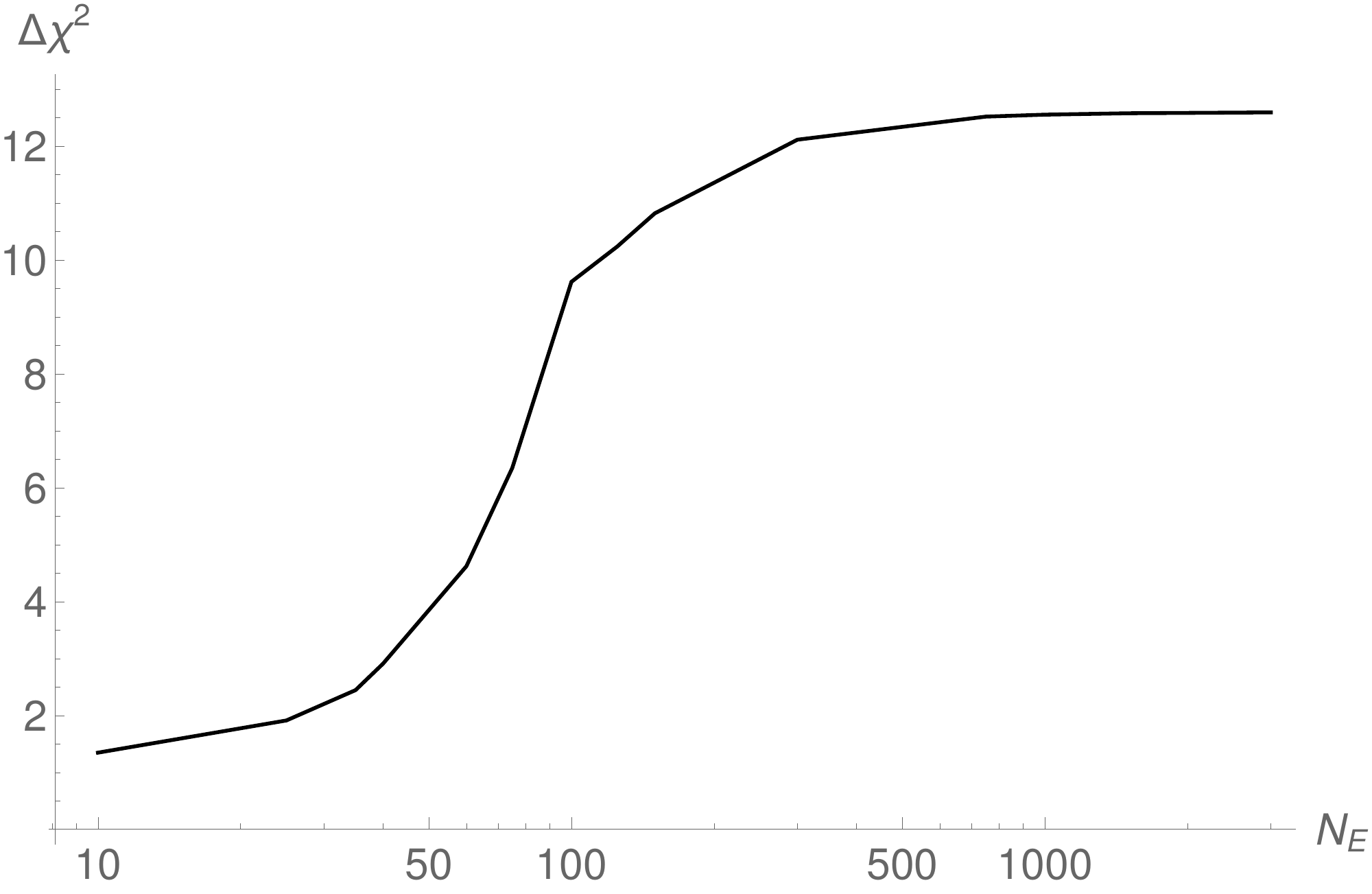}
\end{center}
\caption{\label{SizeBin} $\Delta\chi^2$ as a function of the number of energy bins considered in the 1.5-8.5 MeV region. Only the uncertainty on $\Delta m_{32}^2$ is taken into account.}
\end{figure}
Fig. \ref{SizeBin} shows the expected value of $\Delta\chi^2$ as a function of the number of bins (considering the spectrum in the region between 1.5 and 8.5 MeV); in the following calculations we will use 10 keV energy bins ({\it i.e.} we will consider 700 bins in the said interval). For this reason (\ref{ExpectedEventAnalytical}) can be computed using an discrete approximation, as a product of matrices and vectors; assuming that $E_{i+1}-E_i=\Delta E$ for every $i$,
\begin{equation}\label{ExpectedEvents}
N_{MH,F,i}=\sum_j G_{ij}P'_{MH,j}\sum_K n'_{K,j}
\end{equation}
where here and throughout the paper lower-case Roman indices indicate the energy bin, while upper-case Roman indices indicate the different reactor cores, and
\begin{eqnarray}
&&G_{ij}= G(E_i+\Delta E/2;E_j+\Delta E/2,\sigma^2(E_j+\Delta E/2))\Delta E \nonumber\\ &&P'_{MH,j}=P_{MH}(E_j+\Delta E/2,L) \qquad  n'_{K,j}=\int_{E_j}^{E_{j+1}} \textrm{d}E\phi_K(E)
\end{eqnarray}
At the near detector we can use two approximations to simplify the calculations: first of all we can drop $P'_{MH,j}$, since the neutrino oscillation can be safely neglected, second the detector will receive neutrinos only from one core. We can write
\begin{equation}
N_{N,i}=\frac{1}{\mathcal{R}}\sum_jG_{ij}n'_{K,j}
\end{equation}
where $\mathcal{R}$ is a renormalization factor that takes into account the differences in mass, neutrino flux and baseline between the near and far detector.  We also assume that the far and near detectors have the same energy resolution. Notice that here $K$ is not summed but indicates a specific reactor core.

\subsection{Example: Same chemical composition}
Now we will consider a simple example, where we assume that all the cores have the same chemical composition, hence
\begin{equation}
n'_{R,i}\propto n'_{K,i} \qquad \forall R,K
\end{equation}
Here we have $\propto$ and not $=$ because the reactors can still have different thermal power.

We can express the expected number of events (writing explicitly the dependence on the pull parameters) as
\begin{eqnarray}
N_{F,i}&=\sum_j &G_{ij}P'_{MH,j} (n'_{j}+\beta'_j) \nonumber\\
N_{N,i}&=\sum_j &\frac{1}{\mathcal{R}}G_{ij} (n'_{j}+\beta'_j)
\end{eqnarray}
Here $\mathcal{R}$ is defined as
\begin{equation}\label{DefR}
\mathcal{R}=\frac{L_N^2}{L^2}\frac{ M_F W_F}{M_N W_N }
\end{equation}
where $L$ and $L_N$ are the baselines of the far and near detector, respectively, $M_{N(F)}$ and $W_{N(F)}$ are the mass of the near (far) detector and the thermal power of the core(s) seen by each detector.
Taking into account (\ref{ChiD}), we can rewrite (\ref{ChiTot}) as
\begin{equation}\label{ChiNoChem}
\Delta\chi^2=\chi_{IH}^2=\min_{\beta'_i,\Delta m_{32}^2}\left( \sum_{i}\frac{(\sum_j G_{ij}\Delta P'_{j}n'_{j}-G_{ij} P'_{IH,j} \beta'_{j})^2}{ \sum_j G_{ij}\Delta P'_{NH,j} n'_{j}} + \sum_{i}\frac{ (\sum_j G_{ij}\beta'_{j})^2}{\mathcal{R} \sum_j G_{ij}n'_{j}}  +P.T.\right)
\end{equation}
With perfect energy resolution ({\it i.e.} if $G_{ij}=\delta_{ij}$), the minimization over $\beta'_i$ would be trivial, because the pull parameters in each bin can be minimized separately. However the finite energy resolution mixes all the pull parameters from different energy bins, and in order to obtain an analytical solution it is necessary to compute the inverse of $G_{ij}$, which would lead to huge numerical errors. It is still possible to find the minimum numerically, however due to the large number of parameters it would be quite time-consuming and it would require high computational power. 

It is possible to obtain analytical results by using the following approximation:
\begin{equation}
\sum _j G_{ij}P'_{MH,j} n'_{j}=(\sum _j G_{ij}P'_{MH,j})(\sum_j G_{il}n'_{l})
\end{equation}
The exact results will be sometimes indicated as $(GP'n')$ and the approximation as $(GP')(Gn')$. With this approximation we are performing the convolution with the Gaussian separately for the spectrum and the oscillation probability. Assuming that the energy resolution of the near and far detector is the same, this is equivalent to considering the spectrum seen at the near detector as the ''real'' unoscillated spectrum, and the expected spectrum at the far detector is obtained by multiplying the near detector spectrum with the oscillation probability convoluted with a Gaussian: what is lost in this picture is that, following the exact procedure, in the Gaussian convolution the oscillation probability should be weighted by the value of the spectrum at that particular point, however since the mass hierarchy determination requires a very good energy resolution and the Gaussian must be sharply peaked on the real energy, the error due to this approximation is quite small. Does this mean that this is a reasonable approximation? The answer is ``Not really, but...''. 
\begin{figure}\begin{center}
\includegraphics[width=0.45\textwidth]{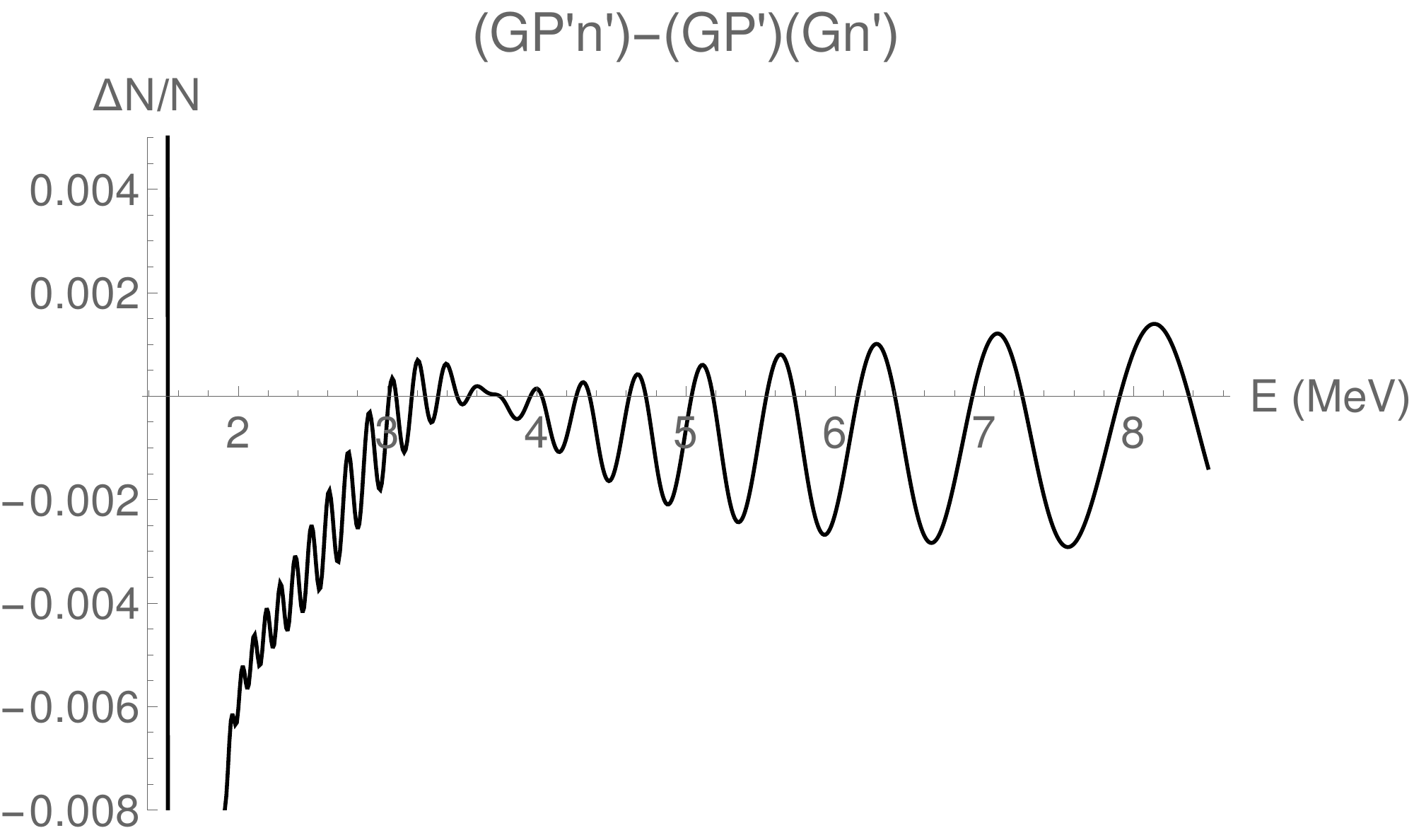}
\end{center}
\caption{\label{DiffAppr} Fractional difference between the spectra obtained with the approximation $(GP')(Gn')$ and with $(GP'n')$}
\end{figure}
As can be seen in Fig. \ref{DiffAppr}, except for the very low energy region (where the spectrum is sharply increasing, this region  does not carry important information for the hierarchy determination), the difference between the exact calculation and the approximation is less than $0.5\%$; however this error is (albeit smaller) of the same order of magnitude as the difference between the hierarchies, this means that it could affect the final result. It is however still possible to use this approximation if it's employed to compute both the fitting model and the Asimov data set: in this way, indeed, the $0.5\%$ error would be not on the spectrum, ({\it i.e.} $GP'n'$), but on the difference between the spectra ($G\Delta P'n'$), which is already very small and it would be a second order effect, that can be neglected. On the other hand, however, this means that this technique can be used only in numerical simulations, while in order to analyze the actual results of an experiment the minimization must be performed numerically. 

In Tab. \ref{TabChi} we show $\Delta\chi^2$ computed in all of the possible cases, assuming a perfect knowledge of the unoscillated spectrum ({\it i.e.} setting $\beta_i=0$) and taking into account only the contribution of the far detector.  We remind the reader that if the approximation is used only for the fit but not for the Asimov dataset, $\chi^2_{NH}$ is not zero anymore, and it must be taken into account. 
\begin{table}[h]
\begin{center}
\begin{tabular}{|c|c|c|} 
\hline Asimov$\backslash$Fit & $(GPn)$ & $(GP)(Gn)$ \\ \hline
  $(GPn)$ & 12.60 & 13.77\\ \hline
$(GP)(Gn)$ & - & 12.53\\ \hline
\end{tabular}
\end{center}
\caption{\label{TabChi}$\Delta\chi^2$ computed with and without the approximation $(GP)(Gn)$}
\end{table}
We can see that if the approximation is used only in the fit, the difference with respect to the exact case is quite large ($\simeq 10\%$), while if $(GP')(Gn')$ is used both for the fit and for the Asimov dataset  the difference is around $0.5\%$. 

We can now rename 
\begin{equation}
P_i=\sum_j G_{ij}P'_j \qquad n_i=\sum_j G_{ij}n'_j \qquad \beta_i=\sum_j G_{ij}\beta'_j
\end{equation}
The $\Delta\chi^2$ now reads
\begin{equation}
\Delta\chi^2=\min_{\beta_i,\Delta m_{32}^2}\left( \sum_{i}\frac{(\Delta P_{i}n_{i}- P_{IH,i} \beta_{i})^2}{P_{NH,i} n_{i}} + \sum_{i}\frac{ \beta_{i}^2}{\mathcal{R} n_{i}}+P.T.\right)
\end{equation}
The minimization over $\beta_i$ is now trivial, and the result is
\begin{equation}
\Delta\chi^2=\chi_{IH}^2=\min_{\Delta m_{32}^2}\left( \sum_j\frac{(\Delta P_i n_i)^2}{P_{NH,i}n_i+\mathcal{R}P_{IH,i}^2n_i}+P.T.\right)
\end{equation}

\section{Chemical Composition}
The neutrino spectrum depends on the chemical composition of the fuel. There were several works in the last few years on this topic, studying, for example, the dependence on the chemical composition of the 5 MeV bump or of the reactor neutrino anomaly \cite{LittlejohnFission,GiuntiFission,DayaBayFission}. Since at JUNO the far detector will receive neutrinos from two power plants (with reactors of different model and generation), while the near detector will see only one core, it is quite likely that the two unoscillated spectra will be different. However, the chemical composition will evolve with time, so at the near detector it will be possible to sample different compositions during the experiment: is it possible to use this information to reconstruct the spectrum at the far detector? We will consider only the contribution of two isotopes to the spectrum $^{239}$Pu and $^{235}$U, that together produce almost 90$\%$ of the neutrinos detected at Daya Bay; increasing this number would complicate the calculations but not give any additional insight on the problem. First of all we will use the same approximation employed in the previous section:
\begin{equation}
(GP'n')\rightarrow (GP')(Gn') \qquad GP'\rightarrow P \qquad Gn'\rightarrow n \qquad G\beta'\rightarrow \beta
\end{equation}
We can write the unoscillated spectrum at the near detector in Eq. (\ref{ExpectedEvents}) as:
\begin{equation}
\sum_{K} n_{K,i}=\rho n_{Pu,i}+(1-\rho)n_{Ur,i}
\end{equation}
where $\rho$ is the fraction of the fissions due to $^{239}$P, averaged over all the cores, and $n_{Pu(Ur),i}$ is the unoscillated spectrum generated by $^{239}$Pu ($^{235}$U); we used ``Ur'' instead of the chemical symbol for Uranium in order to avoid any confusion with the indices used to indicate the reactor cores. 

We must now to introduce two sets of pull parameters $\beta_i$, one for each isotope.  Writing them explicitly we have
\begin{equation}
N_{F,i}=P_{IH,i}\left(\rho (n_{Pu,i}+\beta_{Pu,i})+ (1-\rho)(n_{Ur,i}+\beta_{Ur,i})\right)
\end{equation}
We divide the spectrum at the near detector in $N_t$ time bins, all of the same size; in each time bin $\gamma$, the fraction of fissions due to $^{239}$Pu will be $\rho_\gamma$. In order to be able to reconstruct the unoscillated spectrum at the far detector, $N_t$ must be equal or larger than the number of isotopes. We have
\begin{equation}
n_{N,\gamma,i}=\frac{1}{\mathcal{R}}\left(\rho_\gamma (n_{Pu,i}+\beta_{Pu,i})+ (1-\rho_\gamma)(n_{Ur,i}+\beta_{Ur,i})\right)
\end{equation}
where here and in the rest of the paper the Greek indices ($\gamma$, in this case) indicate the time bins. $\mathcal{R}$ is now defined as
\begin{equation}
\mathcal{R}=N_t\frac{L_N^2}{L^2}\frac{ M_F W_F}{M_N W_N }
\end{equation}
the only difference with respect to Eq. (\ref{DefR}) is the presence of the number of time bins $N_t$. 

We can now write the unoscillated spectrum at the far detector as a linear combination of the spectra measured at the near detector in the different time bins
\begin{equation} \label{EvFarNear}
N_{F,i}=P_{IH,i}\mathcal{R}\sum_\gamma \alpha_\gamma n_{N,\gamma,i}=\left(n_i+\rho\beta_{Pu,i}+(1-\rho)\beta_{Ur,i}\right)
\end{equation}
where $n_i=\rho n_{Pu,i}+ (1-\rho)n_{Ur,i}$ and, in the last step, we imposed the following conditions on the coefficients $\alpha_\gamma$:
\begin{equation}\label{CondAlpha}
\sum_\gamma \alpha_\gamma=1 \qquad \sum_\gamma \alpha_\gamma \rho_\gamma=\rho
\end{equation}
There are two important observations that must be made here: first, if $N_t$ is larger than the number of isotopes considered, the choice of $\alpha_\gamma$ is not unique; however we can notice that $\alpha_\gamma$ do appear not on the right side of Eq. (\ref{EvFarNear}). Indeed, since the $n_i$'s and $\beta_i$'s are multiplied by the same coefficients, the conditions (\ref{CondAlpha}) ensure that $\Delta\chi^2$ does not depends on our choice of $\alpha_\gamma$, which will not affect the final result. Second, it is not always possible to express the expected number of events at the far detector in a simple expression such as Eq. (\ref{EvFarNear}) (for example, this is not possible when the interference is taken into account), however in order to use this approach it is sufficient that we are able to reconstruct the unoscillated spectrum for each isotope as a linear combination of the near detector data. 

The $\Delta\chi^2$ now reads
\begin{eqnarray}
&&\Delta\chi^2=\nonumber \\
&&\min_{\vec{\beta},\Delta m_{32}^2}\sum_i\left(\frac{(\Delta P_i n_i-P_{IH,i}(\rho\beta_{Pu,i}+(1-\rho)\beta_{Ur,i}))^2}{P_{NH,i}n_i} +\sum_\gamma\frac{(\rho_\gamma\beta_{Pu,i}+(1-\rho_\gamma)\beta_{Ur,i})^2}{\mathcal{R}^2 n_{N,\gamma,i}}+P.T.\right)\nonumber\\
\end{eqnarray}
where $\vec{\beta}=(\beta_{Pb,i},\beta_{Ur,i})$. Using the following transformation for the coefficients $\beta$:
\begin{equation}\label{TrasfBeta}
\beta_{Ur,i}\rightarrow\tilde{\beta}_{Ur,i}=(1-\rho)P_{IH,i}\beta_{Ur,i} \qquad \beta_{Pu,i}\rightarrow\tilde{\beta}_{Pu,i}=\rho P_{IH,i}\beta_{Pu,i}
\end{equation}
and expanding the the second term in the $\Delta\chi^2$ we can write:
\begin{equation}
\Delta\chi^2=\min_{\vec{\beta},\Delta m_{32}^2}\sum_i\left(\frac{(\Delta P_i n_i-\tilde{\beta}_{Pu,i}-\tilde{\beta}_{Ur,i}))^2}{\sigma_{Exp,i}^2} +C_{Pu,i}\tilde{\beta}_{Pu,i}^2+C_{Ur,i}\tilde{\beta}_{Ur,i}^2+2C_{Mix,i}\tilde{\beta}_{Ur,i}\tilde{\beta}_{Pu,i}+P.T.\right)
\end{equation}
where
\begin{eqnarray}
&&C_{Pu,i}=\sum_\gamma\frac{\rho_\gamma^2}{\rho^2\mathcal{R}^2P_{IH,i}^2n_{N,\gamma,i}} \qquad C_{Ur,i}=\sum_\gamma\frac{(1-\rho_\gamma)^2}{(1-\rho)^2\mathcal{R}^2P_{IH,i}^2n_{N,\gamma,i}} \nonumber\\
&&C_{Mix,i}=\sum_\gamma\frac{(1-\rho_\gamma)\rho_\gamma}{(1-\rho)\rho\mathcal{R}^2P_{IH,i}^2n_{N,\gamma,i}}
\qquad \sigma_{Exp,i}^2=P_{NH,i}n_i
\end{eqnarray}
 It is now possible to perform analytically the minimization over the $\tilde{\beta}_{Pu,i}$'s and $\tilde{\beta}_{Ur,i}$'s, we have:
\begin{equation}\label{ChiVariTimeBin}
\Delta\chi^2=\min_{\Delta m_{32}^2}\left(\sum_i \frac{(n_i\Delta P_i)^2(C_{Ur,i}C_{Pu,i}-C_{Mix,i}^2)}{(C_{Ur,i}C_{Pu,i}-C_{Mix,i}^2)\sigma_{Exp,i}^2+(C_{Pu,i}+C_{Ur,i}-2C_{Mix,i})}+P.T.\right)
\end{equation}
To obtain these results we have assumed that the energy resolution is the same at the near and far detector; it is worth noticing that, while in the case of identical chemical composition it is possible to get an analytical result for any energy resolution, if the unoscillated spectra at near and far detector are different it is possible to write explicitly the solution only if $\sigma_F>\sigma_N$, otherwise it is necessary to compute the inverse of $G_{ij}$.

Assuming the change of the chemical composition will be constant in time, we have that $\rho_\gamma$ is
\begin{equation}
\rho_\gamma=\rho_{min}+\frac{\rho_{max}-\rho_{min}}{N_t}\left(\gamma-\frac{1}{2}\right)
\end{equation}
where $\gamma=1,\dots,N_t$. We assume that, at the near detector, $\rho_{min}=0.25$ and $\rho_{max}=0.35$ (similar to that reported in \cite{DayaBayFission}, regarding the fuel evolution at Daya Bay; it is worth noticing that the data reported there involve also an average over several cores, so in the case of only one core the variation could be larger). 
\begin{figure}\begin{center}
\includegraphics[width=0.45\textwidth]{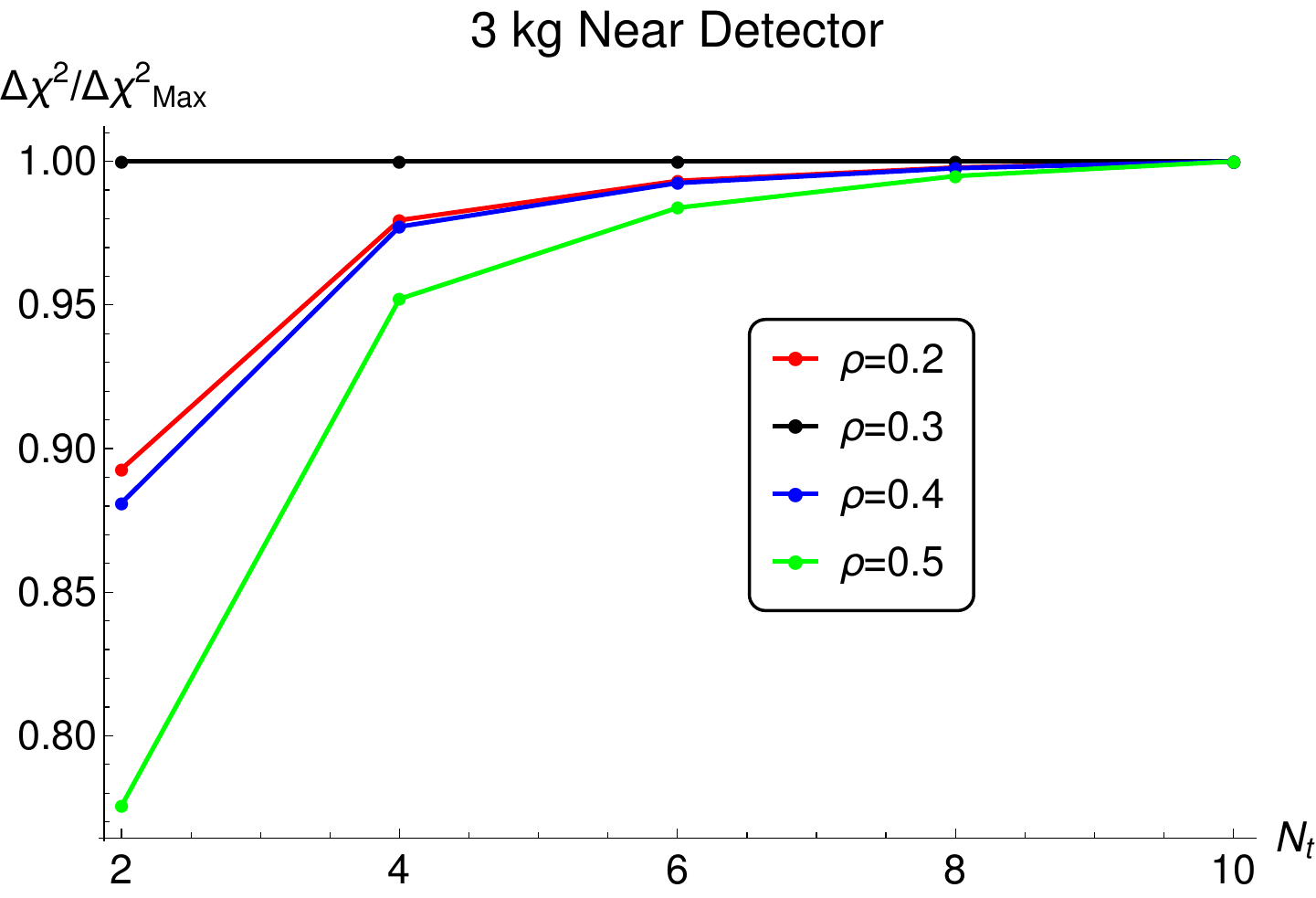}
\includegraphics[width=0.45\textwidth]{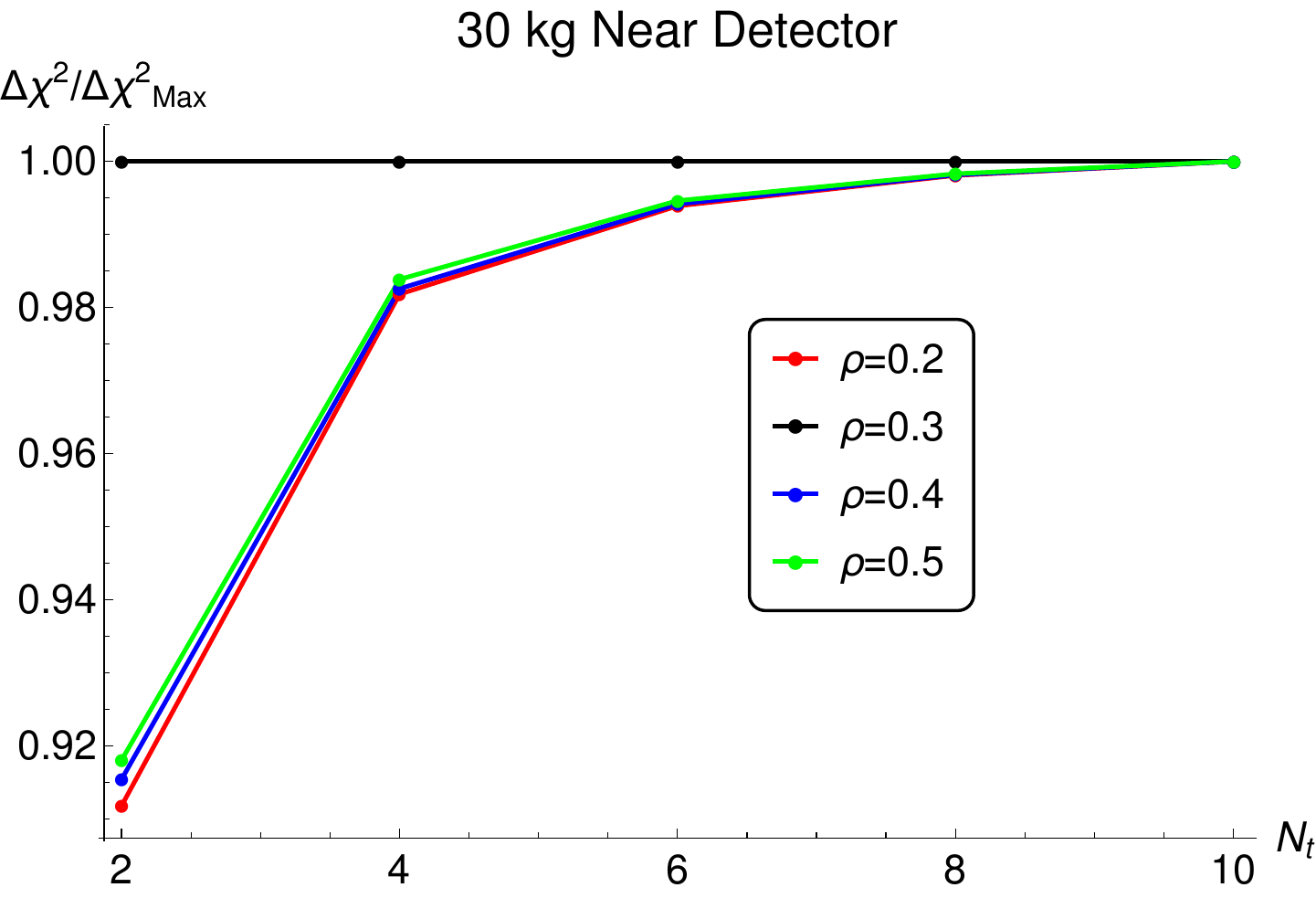}
\includegraphics[width=0.45\textwidth]{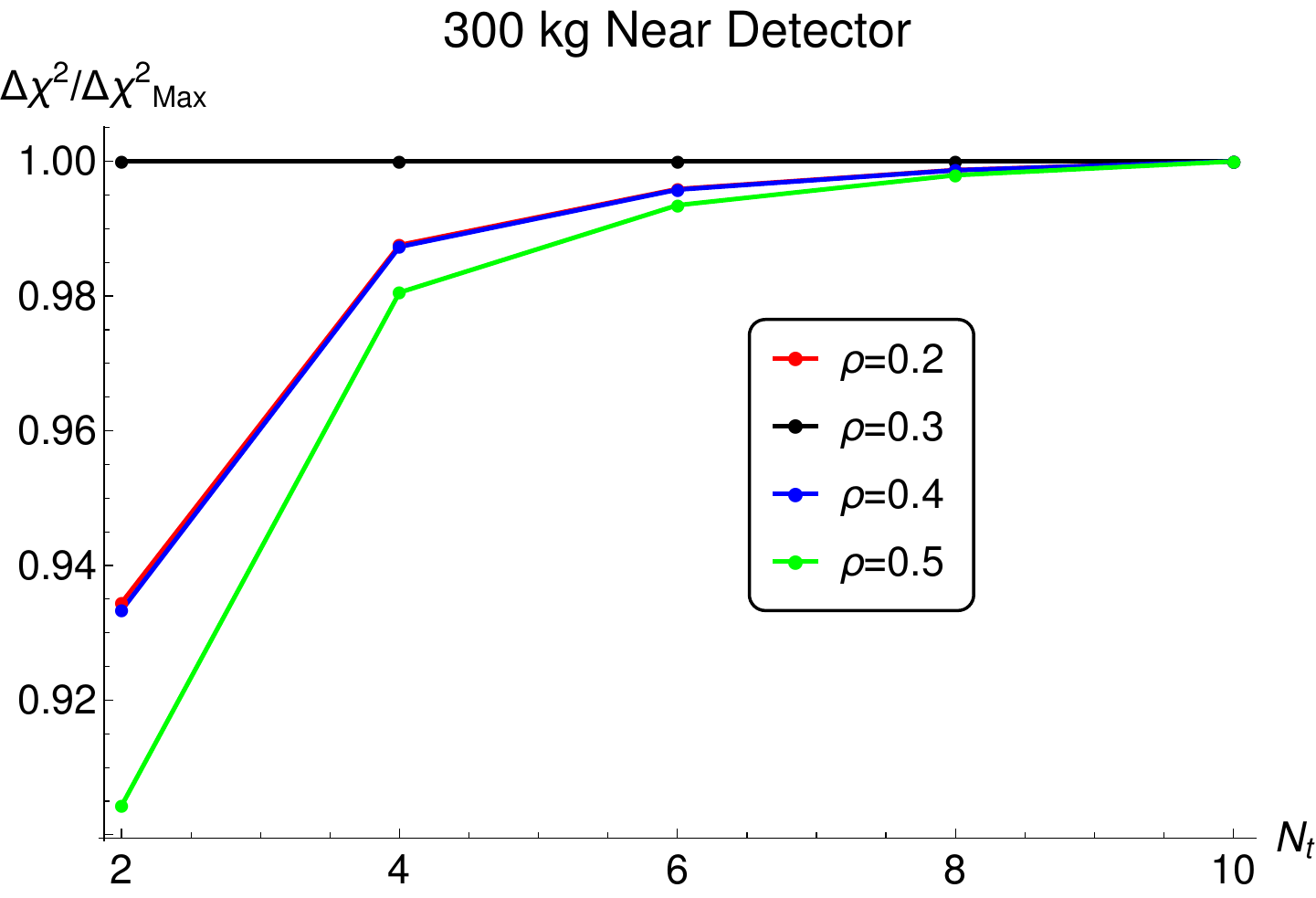}
\includegraphics[width=0.45\textwidth]{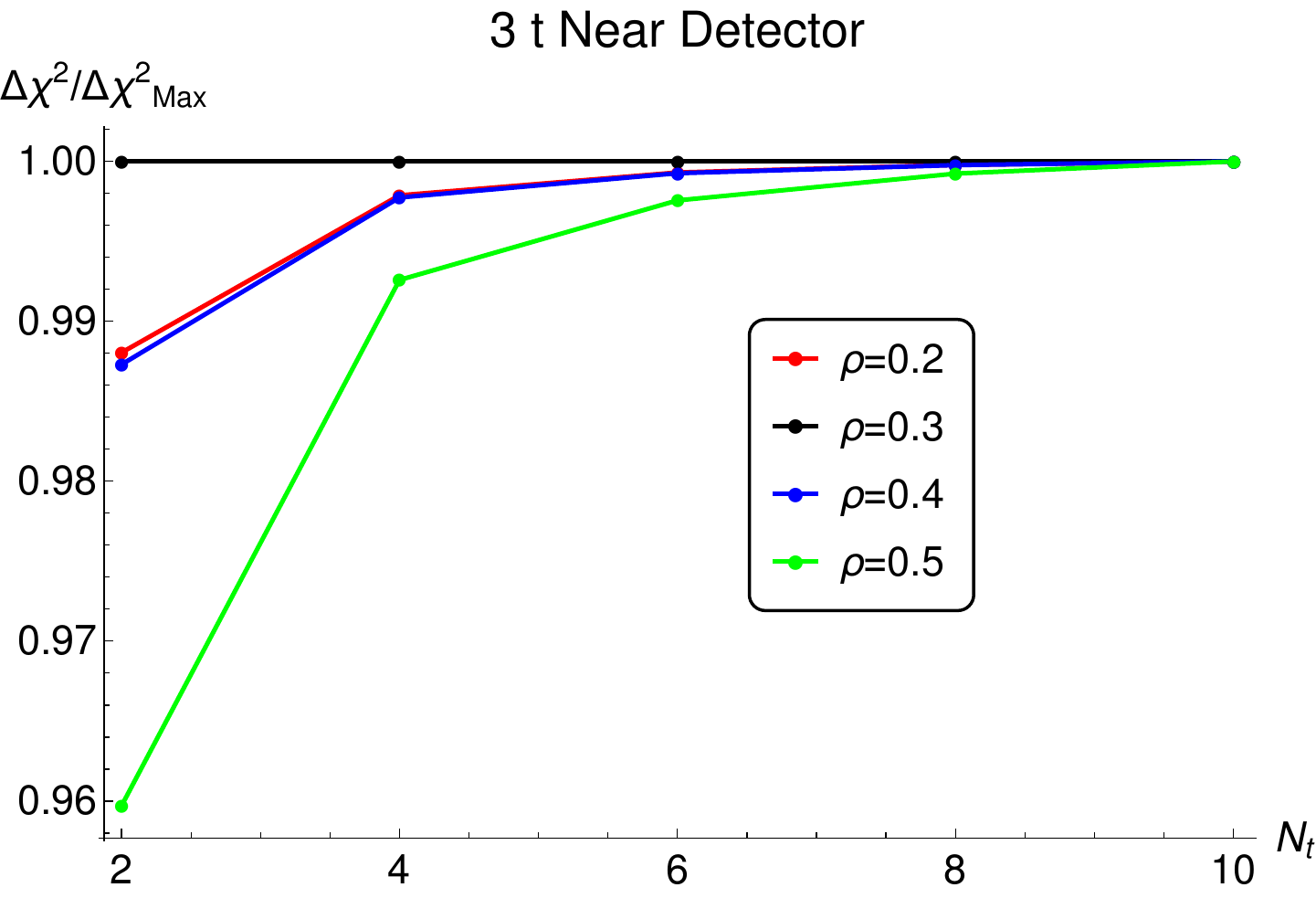}
\end{center}
\caption{\label{ChiTimeBin} $\Delta\chi^2$ as a function of the number of time bins for different masses of the near detector. $\Delta\chi^2_{Max}$ is defined as the $\Delta\chi^2$ with $N_t=10$.}
\end{figure}

\begin{figure}\begin{center}
\includegraphics[width=0.5\textwidth]{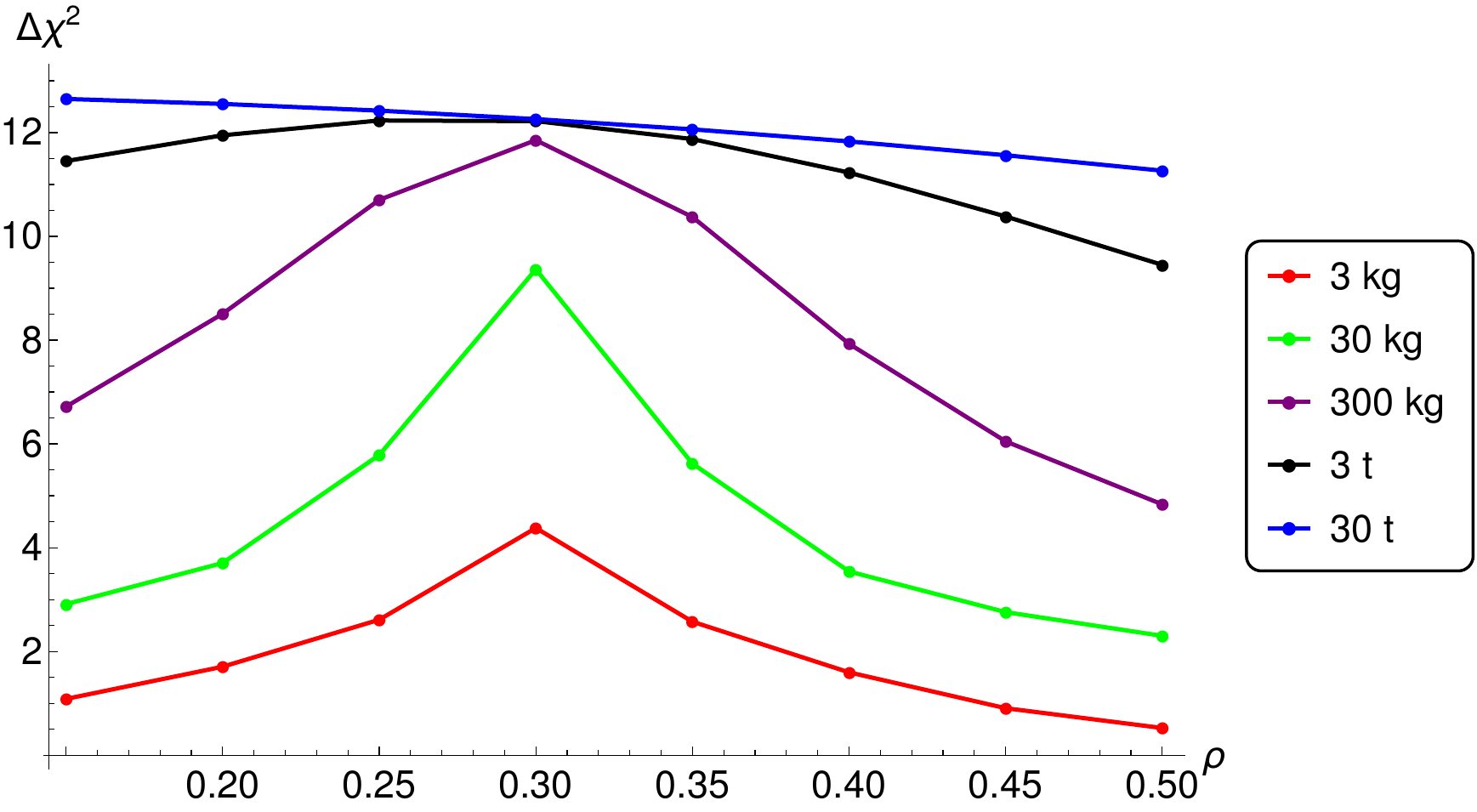}
\end{center}
\caption{\label{ChiChemComp} $\Delta\chi^2$ as a function of the average chemical composition of the far detector spectrum, for different near detector mass.  The peak at $\rho=0.3$ results from the fact that the time-averaged unoscillated spectra at the near and far detectors are proportional.}
\end{figure}

In Fig. \ref{ChiTimeBin} we show $\Delta\chi^2$ as a function of the number of time bins: it is possible to see that increasing $N_t$ will increase the precision, however with a detector in the ton range the difference is minimal. In Fig. \ref{ChiChemComp} we can see the expected value of $\Delta\chi^2$ as a function of the different (average) chemical composition seen by the far detector and of the mass of the near detector; $N_t=2$.

The inclusion of the interference effect is quite straightforward: if the reactor cores are not all at the same distance from the far detector, the expected number of events can be written as
\begin{equation}
N_{F,i}=\sum_K \frac{W_K}{L_K^2}P(L_K)_{NH,i}\left(\rho_K n_{Pu,i}+ (1-\rho_K)n_{Ur,i})\right)
\end{equation}
Here $W_K$ and $L_K$ are the power and the baseline of the $K$-th core, $P(L_K)_{NH,i}$ is the oscillation probability in the $i$-th energy bin, where now the dependence on the baseline is written explicitly. To be consistent with the notation, $n_{Pu,i}$ and $n_{Ur,i}$ must also be adequately normalized, {\it i.e.} they represent the unoscillated spectrum that would be measured at the far detector from a 1 GW core (with the appropriate chemical composition) 1 m away, if the power and the baseline are expressed in GW and m, respectively. 

We can see that now we cannot define a generic ``oscillation probability'', since it will be different for each isotope: we can collect of the factors multiplying the unoscillated spectrum by writing
\begin{eqnarray}
\mathcal{P}_{Pu,MH,i}=\sum_K \frac{\rho_KW_K}{L_K^2}P(L_K)_{MH,i} \qquad \mathcal{P}_{Ur,MH,i}=\sum_K \frac{(1-\rho_K)W_K}{L_K^2}P(L_K)_{MH,i}
\end{eqnarray}
which, of course, do not correspond to an oscillation probability. We can follow the same procedure as before to calculate analytically the minimization over the $\beta$'s, however  Eq. (\ref{TrasfBeta}) now reads
\begin{equation}
\beta_{Ur,i}\rightarrow\tilde{\beta}_{Ur,i}=\mathcal{P}_{Ur,IH,i}\beta_{Ur,i} \qquad \beta_{Pu,i}\rightarrow\tilde{\beta}_{Pu,i}=\mathcal{P}_{Pu,IH,i}\beta_{Pu,i}
\end{equation}
$\Delta\chi^2$ will still have the same expression as in Eq. (\ref{ChiVariTimeBin}), however now the $C$'s and $\sigma_{Exp,i}$ will be defined as
\begin{eqnarray}
&&C_{Pu,i}=\sum_\gamma\frac{\rho_\gamma^2}{\mathcal{R}^2\mathcal{P}_{Pu,IH,i}^2n_{N,\gamma,i}} \qquad C_{Ur,i}=\sum_\gamma\frac{(1-\rho_\gamma)^2}{\mathcal{R}^2\mathcal{P}_{Ur,IH,i}^2n_{N,\gamma,i}} \nonumber\\
&&C_{Mix,i}=\sum_\gamma\frac{(1-\rho_\gamma)\rho_\gamma}{\mathcal{R}^2\mathcal{P}_{Pu,IH,i}\mathcal{P}_{Ur,IH,i}n_{N,\gamma,i}}
\nonumber \\ &&\sigma_{Exp,i}^2=\sum_K \frac{W_K}{L_K^2}P(L_K)_{NH,i}\left(\rho_K n_{Pu,i}+ (1-\rho_K)n_{Ur,i})\right)
\end{eqnarray}

\subsection{Energy Reconstruction}
We conclude this section with an observation: an approach similar to the one described here could be used also for a model-independent treatment of the systematic errors due to the energy non-linearity. We can indeed rewrite Eq. (\ref{ExpectedEventAnalytical}) as
\begin{equation}\label{ExpectedF}
N_{MH,F,i}=\int_{E_i}^{E_{i+1}}\textrm{d}E\int \textrm{d}E' G(E;E',\sigma^2(E'))  P_{MH}(E',L) \sum_K\phi_K(E')=F(E_{i+1})-F(E_i)
\end{equation}
If there is an unknown systematic error in the energy reconstruction, then the visible energy is related to the real energy of the neutrino by the formula
\begin{equation}
E_{obs}(E_{real})=E_{real}(1+\epsilon(E_{real}))
\end{equation}
Eq. (\ref{ExpectedF}) becomes
\begin{equation}
N_{MH,F,i}=F(E_{obs}(E_{i+1}))-F(E_{obs}(E_i))=F(E_{i+1})-F(E_i)+E_i\epsilon_i(F'(E_{i+1})-F'(E_i))+\dots
\end{equation}
where $\epsilon_i=\epsilon(E_i)$ and all the terms proportional to $\epsilon_i(E_{i+1}-E_i)$ are neglected. It is now possible to use the same approach followed in the previous section to take into account the uncertainty due to the non-linearity, without the need to rely on any particular model. It is important to underline, however, that the main difficulty in this procedure is to estimate the penalty term related to $\epsilon(E)$. 

To clarify this point let consider the following example: let us assume that, in the case discussed in the previous section, there is a systematic error in the unoscillated spectrum which is constant for every energy bin, {\it i.e.} $\beta_i=c$. Since the penalty term for $\beta_i$ comes from the measurement at the near detector, and the error in that measurement goes like $1/\sqrt{N_{F,i}}$, it is easy to prove that if we change the size of the bins, the sum of the penalty terms for all the $\beta$'s would remain constant. If, in the case of non-linearity, we assume naively that the uncertainty for each parameter $\epsilon_i$ is $1\%$ (for example), it is also easy to see that, in case of constant bias, the total penalty term would be proportional to the number of energy bins, which is clearly an undesirable feature. This means that the penalty term function could be considerably more complicated, and it could also depend on the choice of the bin size, as well as on the calibration procedure. This also introduces some sort of model-dependence in our analysis (because the final result depends on how the error bars for the non-linearity are chosen), however this is unavoidable, regardless of the method chosen; the main advantage of this approach is that it is not necessary to chose a specific model to parametrize $\epsilon(E)$.

\section{Conclusion}
The uncertainty in the reactor neutrino spectrum could affect the mass hierarchy determination, however we have shown that it is possible to reduce significantly the effect on the final results using the data from a near detector, with an approach that does not rely on any theoretical model; if the mass of the near detector is in the ton range, the loss of precision is negligible. Even if the chemical compositions of the reactors seen by the near and far detector are different, it is possible to reconstruct the unoscillated spectrum at the far detector by studying the time evolution of the near spectrum. One of the consequences is that the data collected at JUNO-TAO could be invaluable not only for JUNO but also for other reactor neutrino experiments, that could use those data, where the spectrum is measured with an excellent energy resolution, to extrapolate the spectrum that will be seen in their detectors, even if the chemical composition of the fuel is different.

\section* {Acknowledgement}

\noindent
EC thanks Bryce Littlejohn and Andrew Conant for the useful discussions and comments. JE is supported by the CAS Key Research Program of Frontier Sciences grant QYZDY-SSW-SLH006 and the NSFC MianShang grants 11875296 and 11675223.  EC is supported by NSFC Grant No. 11605247.  JE and EC also thank the Recruitment Program of High-end Foreign Experts for support.

\end{document}